# Spontaneous Gully Polarized Quantum Hall States in ABA Trilayer Graphene

*F. Winterer[1], A.M. Seiler[2], A. Ghazaryan[3], F.R. Geisenhof[1], K. Watanabe[4], T. Taniguchi[5], M. Serbyn[3], R. T. Weitz[1,2,6,*]*

1. Physics of Nanosystems, Department of Physics, Ludwig-Maximilians-Universität München, Amalienstrasse 54, Munich 80799, Germany
2. 1st Physical Institute, Faculty of Physics, University of Göttingen, Friedrich-Hund-Platz 1, Göttingen 37077, Germany
3. Institute of Science and Technology Austria, Am Campus 1, 3400 Klosterneuburg, Austria
4. Research Center for Functional Materials, National Institute for Materials Science, 1-1 Namiki, Tsukuba 305-0044, Japan
5. International Center for Materials Nanoarchitectonics, National Institute for Materials Science, 1-1 Namiki, Tsukuba 305-0044, Japan
6. Munich Center for Quantum Science and Technology (MCQST), Schellingstrasse 4, Munich 80799, Germany
*thomas.weitz@uni-goettingen.de

**Abstract**

Bernal-stacked multilayer graphene is a versatile platform to explore quantum transport phenomena and interaction physics due to its exceptional tunability via electrostatic gating. For instance, upon applying a perpendicular electric field, its band structure exhibits several off-center Dirac points (so-called Dirac gullies) in each valley. Here, the formation of Dirac gullies and the interaction-induced breakdown of gully coherence is explored via magnetotransport measurements in high-quality Bernal-stacked (ABA) trilayer graphene. In the absence of a magnetic field, multiple Lifshitz transitions as function of electric field and charge carrier density indicating the formation of Dirac gullies are identified. In the quantum Hall regime and high electric fields, the emergence of Dirac gullies is evident as an increase in Landau level degeneracy. When tuning both electric and magnetic fields, electron-electron interactions can be controllably enhanced until the gully degeneracy is eventually lifted. The arising correlated ground state is consistent with a previously predicted nematic phase that spontaneously breaks the rotational gully symmetry.

**Main text**

Since the advent of two-dimensional materials, graphene has attracted widespread attention due to its unusual electronic spectrum hosting massless Dirac fermions [1]. Increasing the number of graphene layers and controlling the stacking order allows for realizing superpositions of massive and massless Dirac fermions as well as chiral quasi-particles with cubic or higher power dispersions [2]. Additionally, the band structure of multilayer graphene can be tuned in-situ by applying electric fields [3,4]. This is especially true for Bernal-stacked trilayer graphene (b-LTG), where it is possible to both access topology transformations of the Fermi surface (Lifshitz transitions [5]) and to explore interaction physics solely via electrostatic gating [6].

In the absence of an electric field, the band structure of b-TLG decomposes into a monolayer-like and a bilayer-like band [7]. An external perpendicular electric field introduces a potential difference $\Delta_1$ between bottom and top layer. As a result, the two bands hybridize, driving the monolayer band to high energy and significantly deforming the low-energy bilayer bands [3,4,8]. The interplay between trigonal warping arising from the skewed interlayer hopping term $\gamma_3$ and the electric field leads to the emergence of two additional sets of three off-center Dirac points related via $C_3$ symmetry, often referred to as Dirac gullies [4,8]. It has been predicted, that upon tuning electric and magnetic fields, electron-electron interactions within the Dirac gullies can be significantly enhanced. As a result, a new nematic ground state emerges that spontaneously breaks the rotational gully symmetry [8]. This correlated state is highly reminiscent of nematic states observed by local scanning-tunneling measurements in Bi [9] and predicted in SnTe [10] crystal surfaces, yet with multilayer graphene offering additional tunability. Recently, quantum capacitance measurements on b-TLG devices have indicated the presence of a correlated ground state at high electric fields [11]. However, transport measurements exploring these correlated states have been elusive so far and it remains unclear, whether interactions merely produce short range ordering or percolating phases [9]. Likewise, Lifshitz transitions are promising candidates for hosting correlated states and have, for instance, been associated with superconductivity [12]. In b-TLG, several Lifshitz transitions driven by the formation of Dirac gullies are within experimental reach. However, they have so far eluded experimental observation in transport measurements, since charge disorder is expected to obscure the electronic signatures of these fragile phenomena.

Here, we report magnetotransport measurements on high-quality b-TLG encapsulated in hexagonal boron nitride with dual graphite gates and graphite contacts (see **Figure 1a** and Supplementary information for fabrication details [13]). The stacking sequence has been verified by comparing electrical measurements at zero and finite magnetic field to theory. **Figure 1b** shows the differential two-terminal resistance $R$ of a h-BN encapsulated b-TLG device with graphite contacts and dual graphite gates as a function of charge carrier density $n$ and electric field $D$ in the absence of a magnetic field. At zero doping, the resistance increases with increasing electric field indicating the opening of a small band gap. This is consistent with tight-binding calculations predicting the emergence of a small band gap of a few meV that saturates within the experimentally accessible electric field range [8,14]. Additionally, the resistance displays a faint discontinuity along a parabola-like contour. This is reminiscent of recent quantum capacitance measurements on b-TLG [11] and transport measurements on Bernal-stacked tetralayer graphene samples [15].

In order to understand the transport features, tight-binding calculations based on the Slonczewski-Weiss-McClure [16] model with parameters as inferred in Ref. [11] to simulate the band structure of b-TLG have been conducted. **Figure 1c** shows the calculated density of states (DOS) as function of charge carrier density and electric field, which can be naturally explained via the evolution of the band structure. **Figure 1d** shows the band structure of b-TLG with the emergent Dirac gullies at a potential difference between adjacent graphene layers of $\Delta_1$ = 50 meV. The potential difference $\Delta_1$ is directly proportional to the electric field and can be controlled experimentally by electrostatic gating. When increasing the electric field and, thus, $\Delta_1$, the Dirac gullies move further apart from each other in momentum space (see **Figure 1e-g**). In fact, simulations show that the inter-gully distance $\Delta k$ (see **Figure 1d**) depends approximately linearly on the electric field $\Delta k \sim D \sim \Delta_1$ [8]. At low doping and high electric fields, the Fermi surface is composed of three disconnected pockets, whereas at high doping the Fermi surface consists of a single connected surface. In-between, when tuning the charge carrier density, the Fermi surface undergoes multiple discontinuous changes of its topology [5]. These Lifshitz transitions [5] are accompanied by a logarithmically diverging DOS leading to anomalies in the conductance [8,15]. Comparing the conductance map in **Figure 1b** to the predicted DOS in **Figure 1c** yields fair qualitative agreement. Yet, the transport features are much fainter than expected and are suggestive of a weaker increase in the DOS than predicted. In addition, no signatures of correlated states as a consequence of a diverging DOS have been observed in agreement with previous measurements [11].

After investigation of transport without magnetic field, we turn to study the properties of the b-TLG subject to perpendicular magnetic field. The reorganization of the Fermi surface with the emergence of Dirac gullies has pronounced effect on the structure of Landau levels (LL) and their degeneracy in the quantum Hall regime. **Figure 2a,b** shows the derivative of the conductance as a function of charge carrier density and electric field for two different magnetic fields $B$ = 0.7 T and $B$ = 1.25 T respectively (a zoom-in on the measurement data of **Figure 2a** is shown in **Figure S5** in the supporting information). Here, regions with vanishing $dG/dn$ indicate conductance plateaus that are associated with the formation of LLs. In the absence of an electric field, the most prominent conductance plateaus appear at filling factors $\nu =$ 2, (4), ±6, ±10 (see **Figure 2c,d**) consistent with previous studies [11,17,18]. When increasing the electric field, a complex landscape of LL crossings emerges with pronounced plateaus at filling factors $\nu$ = -3, ±6, 9, ±12 becoming apparent at high electric field (see **Figure 2c,d**).

In order to understand these findings, we have simulated the evolution of LLs with respect to the potential difference $\Delta_1 \sim D$ and the magnetic field $B$ in a single-particle framework. The energies of the spin-degenerate LLs as a function of $\Delta_1$ and $B$ are shown in **Figure 2e,f.** The resulting DOS in the quantum Hall regime (after artificial broadening of the LLs) with respect to the charge carrier density $n$ and potential difference $\Delta_1$ are shown together with the measurement data in **Figure 2a,b**. Without an electric field, in agreement with our measurement data, LLs are spin degenerate and feature nearly perfect valley degeneracy except for the zero energy LLs where valley degeneracy is broken [7,17,19].

At high electric fields, the pattern of LL degeneracy is strongly reorganized: The formation of three Dirac gullies in the vicinity of the K and K' point leads to the formation of six-fold degenerate LL triplets from three intertwining LLs for electrons and holes each (including a factor of 2 coming from the additional spin-degeneracy). Each of these quasi-degenerate triplets are associated with one particular set of $C_3$-related gullies. The triplet wave functions inherit the $C_3$ symmetry and consist of a coherent superposition of contributions from all three gullies [8]. This is consistent with our measurement data, yet with broken spin degeneracy. Comparing the LL crossings observed experimentally to the single-particle simulation (see **Figure 2a,b**) shows excellent agreement. Additional measurements at higher charge carrier densities underline this close correspondence further and are presented in Figure S3 in the supporting information. By matching the positions of prominent LL crossings, $\Delta_1$ can be related to $D$ via an empiric conversion factor $\gamma = 0.073$ /nm assuming a linear relationship $\Delta_1 = \gamma D$.

It is also worth noting that the value of electric field where LLs enter the triplet plateau regime increases with the magnetic field: while the prominent $\nu$ = -3 plateau is already visible at $D \approx 170$ mV nm$^{-1}$ for $B$ = 0.7 T, it emerges only above $D \approx 260$ mV nm$^{-1}$ for $B$ = 1.25 T. This observation aligns well with the single-particle simulations that predict a significantly higher energy splitting of the triplet states for larger magnetic fields and can be understood as follows. At low magnetic fields and high electric fields, inter-gully tunneling is weak and the triplet states are quasi-degenerate [8,20]. Upon increasing the magnetic field, inter-gully tunneling becomes more and more dominant and the triplet state energy splitting increases until the triplets entirely lose their gully character [8,20]. This magnetic breakdown occurs when the magnetic length $l_B = \sqrt{\hbar/(eB)}$ with $e$ being elementary charge becomes comparable to the distance $\Delta k$ between Dirac gullies in $k$-space ($l_B \Delta k \approx 1$) [8]. Since $\Delta k \sim D$ as discussed earlier and shown in **Figure 1e-g**, both high electric fields and low magnetic fields stabilize gully triplets [8]. Thus, the critical electric field $D_C$ below which gully physics breaks down and the triplet state energy splitting becomes significant can be related to the magnetic field via $D_C \sim \sqrt{B}$ [8]. This relation indeed agrees well with the measurement data (see **Figure 2a,b**).

In order to investigate the formation of triplet states further, we show the measurements of the differential conductance $dG/d\nu$ as a function of the filling factor $\nu$ and the magnetic field at fixed value of $D$ = 0 mV nm$^{-1}$ (a) and $D$ = 0.8 V nm$^{-1}$ in **Figure 3a-b** (see also **Figure S4** of the supporting information for data at an intermediate displacement field $D$ = 0.4 mV nm$^{-1}$). In agreement with the observations discussed above and theoretical simulations shown in **Figure 3c,d**, plateaus associated with gully triplets stabilize at low magnetic fields and high electric fields.

While the aforementioned features of the magnetotransport measurements are consistent with the single-particle simulations, many observations cannot be explained in a non-interacting framework. For instance, in the absence of electric fields, all plateaus at integer filling factors between -10 ≤ $\nu$ ≤ -2 are already well-resolved at $B$ = 1.25 T, with some plateaus being visible even well below $B$ = 0.7 T (**Figure 3a**). The observation of breaking of spin and valley degeneracy due to electron-electron interaction at such low magnetic fields further confirms the high quality of the investigated flake [6,18]. Furthermore, in the gully regime at high electric and low magnetic fields, multiple integer plateaus at filling factors -12 ≤ $\nu$ ≤ -12 are well resolved. This is especially apparent for the $\nu$ = 6 and $\nu$ = 12 triplet: Below a critical electric field and above a critical magnetic field, several LL crossings stemming from the intertwining triplet states are visible. When going across the critical field, LL crossings are completely absent and all integer LLs are readily observable (highlighted by the green box in **Figure 2a,b** and **Figure 3a,b**). This indicates that both spin and gully degeneracy are broken. In contrast, the gully degeneracy of the $\nu$ = - 6 triplet seems to be much more persistent and only spin degeneracy appears to be broken. However, at maximum electric field and $B$ = 0.7 T, first indication of developing integer plateaus of the $\nu$ = -6 triplet are evident (see **Figure 2c**). Together, these observations strongly hint towards an interaction-induced symmetry-breaking of triplet states. Indeed, in a simplistic picture, increasing the electric field quenches the energy splitting of the individual LLs that comprise the triplet states. Thus, the Coulomb interaction energy $E_C = e^2/(\varepsilon_{hBN} l_B)$, although being screened by the hBN ($\varepsilon_{hBN} \approx 6.9$ [21]), eventually becomes dominant compared to the kinetic energy and interaction physics is expected to prevail [11]. First indication of a broken triplet degeneracy has also been seen in previous capacitance measurements [11], yet neither its transport properties nor the ordering length scale could be accessed so far.

In order to understand this breakdown of the single-particle picture in more detail, a variational Hartree-Fock analysis to model interaction effects and to reveal the ordering of the ground state of the system has been employed [8] where only the lowest energy state (i.e. 1/3 filling of a triplet ignoring spin) was considered. For small electric fields within the gully regime, the Hartree-Fock ground state coincides with the single-particle ground state. The wave function is completely gully coherent and obeys $C_3$ symmetry. At high electric fields, however, the analysis yields that $C_3$ symmetry is spontaneously broken and the ground state is a nematic gully polarized state [8]. The transition between these two regimes is of the first order and takes place at a critical electric field $D_C^{\mathrm{HF}}$. This is consistent with the measurement results that display a sudden change in LL degeneracy that is associated with a critical field (green box in **Figure 2a,b** and **Figure 3a,b**). Furthermore, the measurements also show a higher critical field for the $\nu$ = 12 triplet compared the low energy $\nu$ = 6 triplet in agreement with the Hartree-Fock simulations [8]. Although these measurements strongly indicate the presence of a nematic phase, further measurements of transport anisotropies are needed to verify the symmetry of the correlated ground state.

In conclusion, using h-BN encapsulated, dually gated b-TLG we have electrostatically tuned both the electric field and the Fermi energy and observed the Lifshitz transitions necessitated by the formation of Dirac gullies in transport. In the quantum Hall regime, the Dirac gullies give rise to the formation of triplet states that modify the sequence of LLs. By changing electric and magnetic fields, we tuned the relative importance of electron-electron interactions and use transport measurements to reveal the emergence of a interaction-driven ground state that breaks the triplet state degeneracy. This correlated state is consistent with Hartree-Fock simulations predicting a gully polarized nematic ground state at high electric fields that spontaneously breaks rotational symmetry.

**Acknowledgements**

We acknowledge funding from the Center for Nanoscience (CeNS) and by the Deutsche Forschungsgemeinschaft (DFG, German Research Foundation) under Germany's Excellence Strategy-EXC-2111-390814868 (MCQST). K.W. and T.T. acknowledge support from the Elemental Strategy Initiative conducted by the MEXT, Japan (Grant Number JPMXP0112101001) and JSPS KAKENHI (Grant Numbers 19H05790 and JP20H00354).

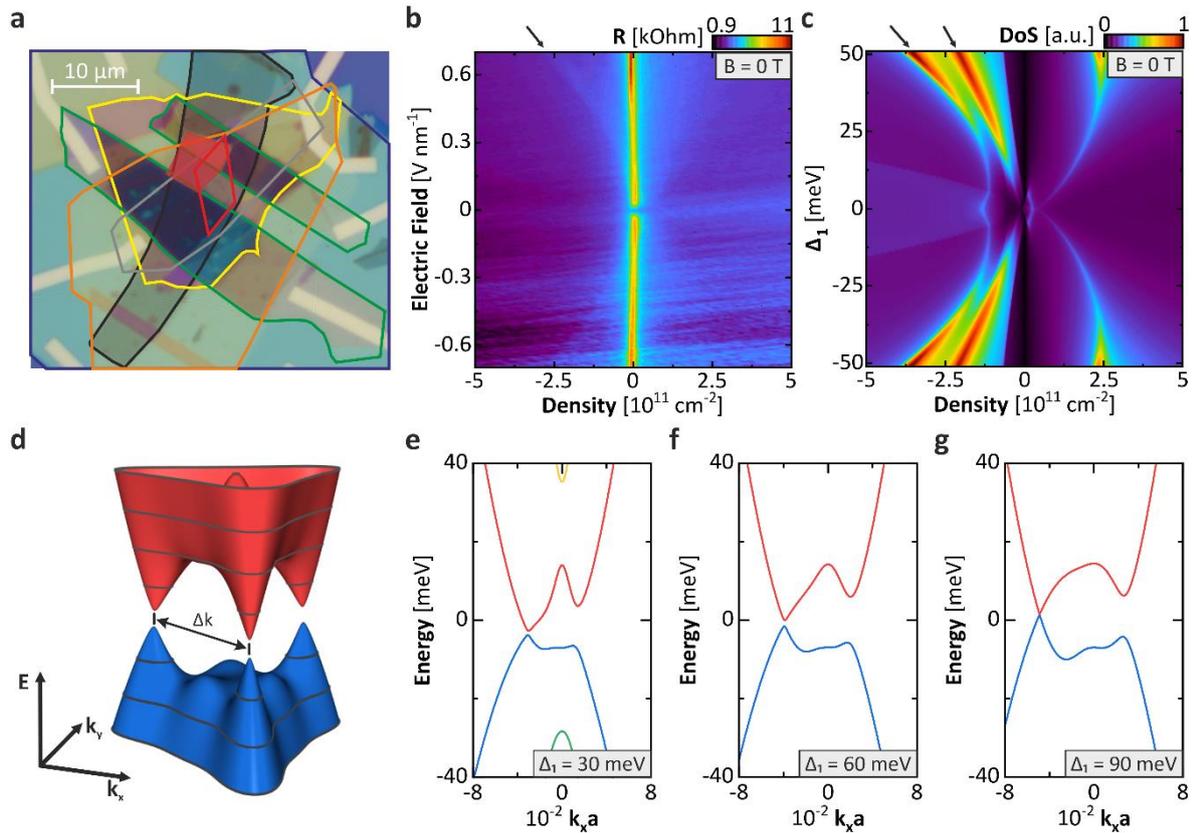

**Figure 1: Device Geometry and Electronic Properties of b-TLG. (a)** Microscope image of the sample geometry consisting of the following layers: bottom graphite gate (black), bottom hBN (blue), trilayer graphene (red), graphite contacts (green), top hBN (orange), top graphite gate (gray) and top hBN (yellow). The red shaded area has been etched prior to the measurements. **(b)** Differential resistance in logarithmic color scale. The drift in graphite contact resistance with electric field has been corrected to enhance contrast. The arrow indicates a distinct parabola-shaped discontinuity that is associated with a Lifshitz transition. **(c)** Simulation of the DOS as a function of charge carrier density and potential difference $\Delta_1$. At parabolic contours, the Fermi surface topology changes (Lifshitz transition) leading to regions with an exceptionally high DOS. The scale of $\Delta_1$ matches the scale of the electric field in (b). **(d)** Simulated three-dimensional band structure of b-TLG at $\Delta_1 = 50$ meV demonstrating the emergence of Dirac gullies. The distance between these gullies is often referred to as $\Delta k$. **(e-g)** Evolution of the band structure of b-TLG for potential differences 30 meV (e), 60 meV (f) and 90 meV (g). Increasing $\Delta_1$ drives the monolayer bands (green and yellow) to higher fields and leads to an increase in inter-gully distance $\Delta k$.

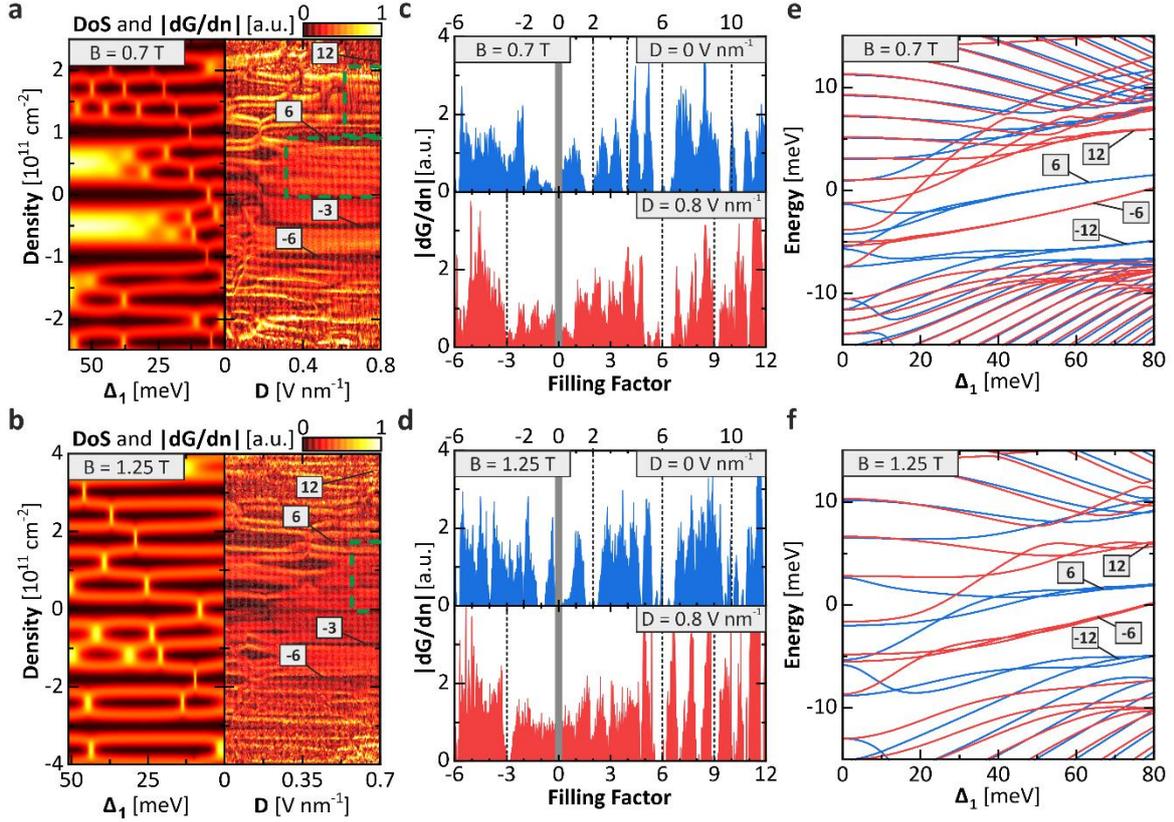

**Figure 2 Landau Levels in Electric Fields: (a,b)** Simulated DOS (left) and measured derivative of the conductance with respect to the charge carrier density (right) of b-TLG as a function of potential difference $\Delta_1$ and electric field $D$ at $B$ = 0.7 T (a) and $B$ = 1.25 T (b). Dark regions correspond to vanishing DOS or derivative of the conductance (i.e. a conductance plateau) respectively. The region of full triplet state splitting is highlighted in green. **(c,d)** Line cuts along electric fields of $D$ = 0 V/nm (blue) and $D$ = 0.8 V/nm (red) at $B$ = 0.7 T (c) and $B$ = 1.25 T (f). The dotted lines indicate the sequence of prominent plateaus at multiples of two (c) and three (d). **(e,f)** Simulated evolution of spin-degenerate LLs stemming from the $K$ (red) and $K'$ (blue) valley as a function of the potential difference $\Delta_1$ at $B$ = 0.7 T (d) and $B$ = 1.25 T (e). The triplet states are denoted according to their associated filling factors $\nu$ and are also highlighted in (a,b). The charge neutrality point is located between the $\nu$ = 6 and $\nu$ = -6 triplets.

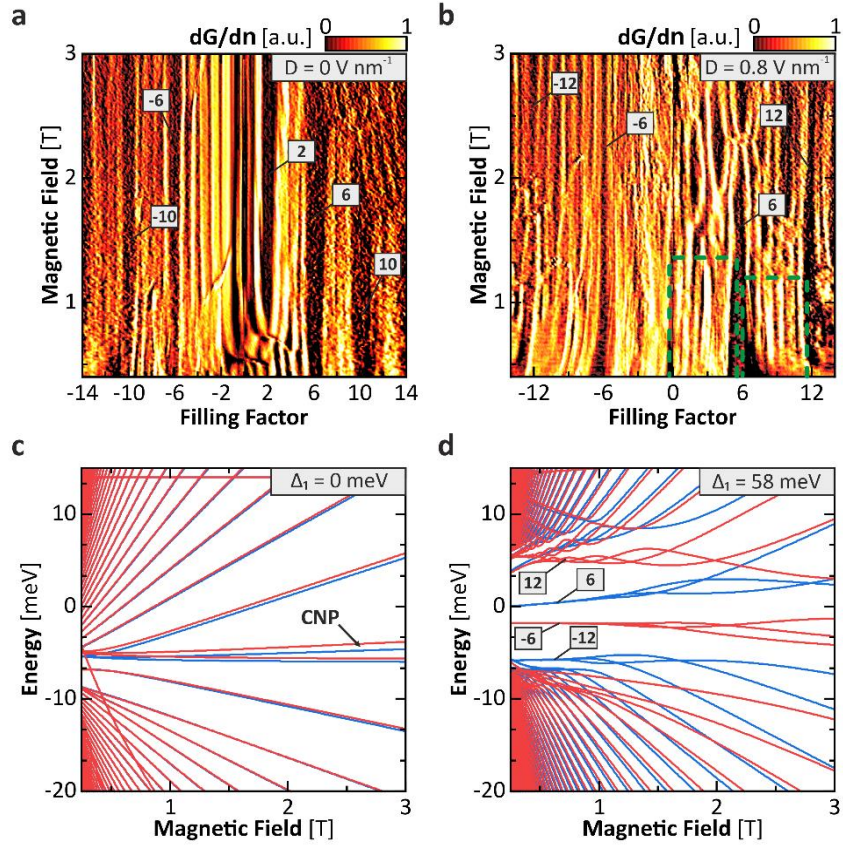

**Figure 3 Landau Levels as a function of Magnetic Fields: (a,b)** Derivative of the conductance with respect to the filling factor at $D = 0$ mV nm$^{-1}$ (a) and $D = 0.8$ V nm$^{-1}$ (b). Dark regions correspond to vanishing derivative of the conductance (i.e., a conductance plateau). For low magnetic and high electric fields (highlighted in green), the degeneracy of several gully triplets is fully split. **(c,d)** Simulated evolution of spin-degenerate LLs at potential differences of $\Delta_1 = 0$ meV (c) and $\Delta_1 = 58$ meV (d) stemming from the $K$ (red) and $K'$ (blue) valley as a function of the magnetic field. The values for $\Delta_1$ in (c) and (d) correspond to the electric fields in (a) and (b) respectively. The triplet states are denoted by their associated filling factor $v$. The charge neutrality point (CNP) is located between the $v = 6$ and $v = -6$ triplets

Supplemental Material

for

# Spontaneous Gully Polarized Quantum Hall States in ABA Trilayer Graphene

F. Winterer[1], A.M. Seiler[2], A. Ghazaryan[3], F.R. Geisenhof[1], K. Watanabe[4], T. Taniguchi[5], M. Serbyn[3], R. T. Weitz[1,2,6,*]

1. Physics of Nanosystems, Department of Physics, Ludwig-Maximilians-Universität München, Amalienstrasse 54, Munich 80799, Germany
2. 1st Physical Institute, Faculty of Physics, University of Göttingen, Friedrich-Hund-Platz 1, Göttingen 37077, Germany
3. Institute of Science and Technology Austria, Am Campus 1, 3400 Klosterneuburg, Austria
4. Research Center for Functional Materials, National Institute for Materials Science, 1-1 Namiki, Tsukuba 305-0044, Japan
5. International Center for Materials Nanoarchitectonics, National Institute for Materials Science, 1-1 Namiki, Tsukuba 305-0044, Japan
6. Munich Center for Quantum Science and Technology (MCQST), Schellingstrasse 4, Munich 80799, Germany
*thomas.weitz@uni-goettingen.de

## Fabrication

In order to fabricate encapsulated graphene devices, a van-der-Waals transfer technique that avoids interlayer contamination was adapted from P. J. Zomer [1] and D. G. Purdie [2]. In this method, flakes are transferred by means of a stamp that consists of a block of polydimethylsiloxane (PDMS) and a film of polycarbonate (PC) [3]. To build the stamp, PDMS was prepared using a commercially available kit (Sylgard 184, Dowsil) with a mass ratio of 10.5:1 between base and curing agent. After preparation, PDMS films with a thickness of approximately 2 mm were cast and dried in vacuum for 24 h to remove bubbles. PC films with a thickness of approximately 10 – 20 µm were prepared using a film application machine (ZAA 2300, Zehntner) with an 8 wt.% solution of PC (Poly(Bisphenol A carbonate), Sigma-Aldrich) in chloroform. Afterwards, the PC films were dried for at least 1 h in air. Having prepared all components, a block of 2 × 2 mm of PDMS was cut out of the PDMS film and placed onto a precleaned glass slide. A window slightly larger than the PDMS block was cut into a double-sided adhesive tape and placed onto the glass slide, such that the PDMS block sat inside the window of the adhesive tape. The PC film was picked up with another tape (again with a window slightly larger than the PDMS block) and placed on top of the PDMS block.

Trilayer graphene flakes are exfoliated and identified by optical microscopy, Raman spectroscopy and atomic force microscopy (AFM). Subsequently, a homogeneous region was cut out using electrode-free scanning probe lithography [3]. Hexagonal boron nitride (hBN) (synthesized as described previously) [4]and graphite flakes are obtained in the same way and their quality is ensured using optical microscopy and AFM. Low-ohmic electrical contact to the flake is established with two additional single-crystal graphite flakes in order to enable the use of a high-quality graphite top gate.

The process of building a stack of encapsulated graphene is shown schematically in **Figure S1**. First, the top hBN flake was picked up. This was done by heating the sample stage to 40 °C and bringing the stamp into contact with the substrate. By heating the sample stage over the course of 20 min to 60 °C, the contact area between stamp and substrate increases and brings the hBN in contact with the stamp. Cooling down the stage back to 40 °C slowly retracts the stamp and delaminates the hBN from the substrate. It is worth noting, that these elevated temperatures are chosen to increase the adhesion of hBN to PC relative to $SiO_2$[2]. This process was repeated in order to pick up graphite contacts, the multilayer graphene flake, the bottom hBN flake and, lastly, the bottom graphite gate. In this process, the strong adhesion of graphene to hBN facilitates the delamination of flakes from the $SiO_2$. Except the first hBN flake, contact between the stamp and subsequent layers of the stack is minimized greatly reducing interlayer contamination. Once all flakes had been stacked onto each other, a clean substrate was prepared and heated to 180 °C above glass transition temperature of PC ($T_G \approx 150$ °C [2,5]). Then, the stamp was tilted by a few degrees and brought into contact with the substrate. By slowly advancing the contact front, trapped interlayer contaminants can be pushed out yielding large, blister-free areas.

At 180 °C, the PC primarily adheres to the SiO$_2$ instead of the PDMS. Thus, after 30 mins, the stamp can slowly be peeled away leaving the stamp together with the PC melted onto the substrate. In a last step, the PC was removed by soaking the sample in chloroform for at least 1 h. Thereafter, the stack was annealed in a vacuum chamber at a pressure of less than $1 \times 10^{-8}$ mbar and 200 °C for 12 h to reduce the number of bubbles further. The top gate was added by repeating this process: First, an hBN and, subsequently, a graphite flake was picked up. Then, both flakes were melted onto the already prepared stack. Finally, the PC was removed with chloroform and the stack was annealed again.

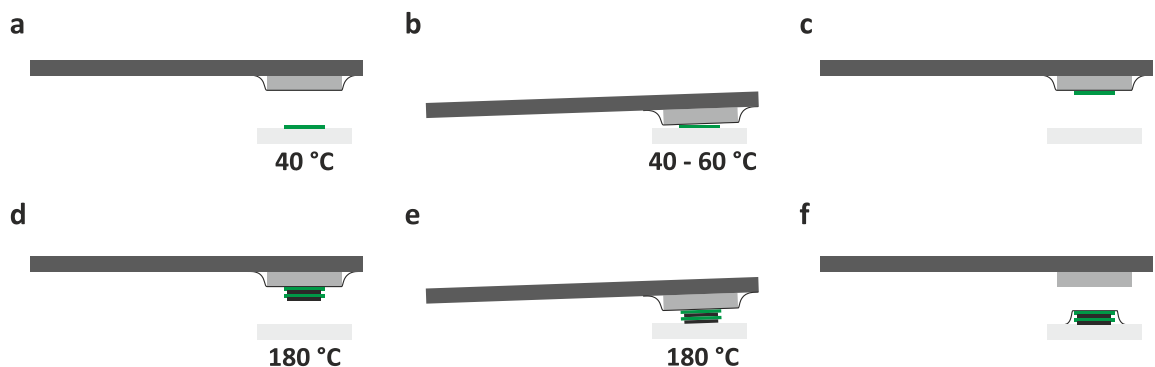

**Figure S1 Flake Transfer: (a)** First, a hBN flake forming the basis of the stack has to be picked up. The hBN flake was positioned directly below the stamp and the substrate was heated to 40 °C. **(b)** In order to pick up the hBN flake, the stamp was slowly brought into contact with the substrate. Then the temperature was ramped to 60 °C and back down to 40 °C to bring the PC in contact with the flake during heating and delaminate it from the substrate upon cooling. **(c)** After pick-up, the stamp with the hBN flake was retracted. **(d)** Once all desired flakes had been picked up using the technique shown in (a-c), the substrate was heated to 180 °C. **(e)** The stack was laid down by slowly bringing the stamp and the stack into contact with the substrate. **(f)** The PC was melted together with the stack onto the substrate and the stamp was retracted.

# Calibration of Dual-Gate Transport

All measurements are done at cryogenic temperatures of < 10 mK in a dilution refrigerator using standard lock-in measurement technique at an AC frequency of 78 Hz and currents of ≤ 20 nA. The stacking order is verified by comparing the Lifshitz transitions at zero field and Landau level crossings in the quantum Hall regime to theory (see **Figure 1** and **Figure 2**). In fact, the stacking order was found to have transformed from ABC to ABA during the transfer process [6]. In our dually gated samples, the charge carrier density $n = C_B(\alpha V_T + V_B)/e$ and the electric field $D = C_B(\alpha V_T - V_B)/2\varepsilon_0$ can be tuned independently as a function of bottom gate voltage $V_B$ and top gate voltage $V_T$. Here, $C_B$ is the capacitance per unit area of the bottom gate and $\alpha = C_T/C_B$ is the ratio between the capacitances of bottom and top gate. Inverting these equations yields $V_B = (en - 2\varepsilon_0 D)/2C_B + V_{B,0}$ and $V_T = (en + 2\varepsilon_0 D)/2\alpha C_B + V_{T,0})$, where the voltage offsets $V_{B,0}$ and $V_{T,0}$ have be introduced to account for residual contaminants that shift the charge neutrality. To determine $\alpha$, the conductance of the device is mapped as a function of the bottom gate voltage and top gate voltage at zero magnetic field as shown in **Figure S2a**. In every line of constant top gate voltage, the center of the conductance dip is determined via a Gaussian fit. Collecting all peak positions and performing a linear fit directly yields a value for $\alpha$ as the inverse slope. This line fit is essentially the charge neutrality line and moving along this line translates to changing the electric field only. Since the resistance has a minimum for vanishing electric field, the offsets $V_{B,0}$ and $V_{T,0}$ are determined by measuring the resistance with respect to bottom gate voltage and top gate voltage along the charge neutrality line (see **Figure S2b,c**). The contact resistance and the gate capacitance of the bottom gate $C_B$ are determined using the quantum Hall effect. A first estimate of $C_B$ is already given by geometrical considerations. Matching the position of conductance plateau to the theoretically expected values of $n = \nu eB/h$ yields the effective $C_B$. The contact resistance $R_C$ can then be determined by matching the conductance $G$ of the plateaus to the expected values of $G = \nu e^2/h$.

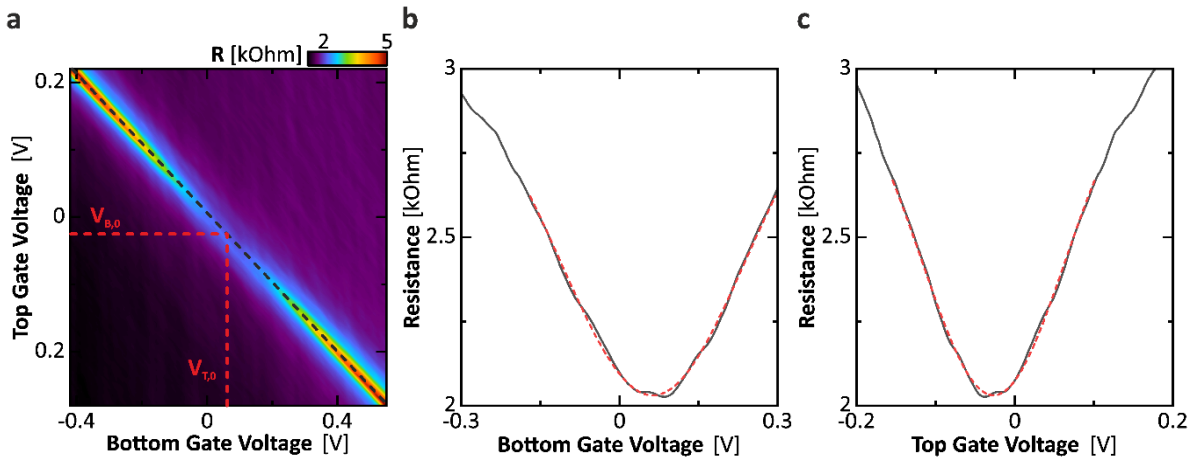

**Figure S2 Calibration of Charge Carrier Density and Electric Field: (a)** Resistance as a function of the bottom gate and the top gate voltage. On every column of the plot, the resistance peak position is extracted by performing a Gaussian fit. Connecting all peak positions and fitting a line yield the charge neutrality line (dotted black). **(b,c)** Resistance as a function of bottom gate voltage (b) and top gate voltage (c) measured along the charge neutrality line. By fitting a Gaussian to the resistance minimum of the line profile with respect to the bottom gate and top gate voltage, the offset voltages $V_{B,0}$ and $V_{T,0}$ can be extracted. These offset voltages are also indicated in (a).

## Magnetotransport at High Charge Carrier Densities

**Figure S3** shows the derivative of the conductance with respect to charge carrier density and electric field for different magnetic fields together with the corresponding single-particle simulations in an extended density range. In the high-density regime, a characteristic sequence of Landau level crossings along parabola-like contours is apparent. From the low charge carrier density measurements (see **Figure 2a,b** in the main text), the interlayer potential difference $\Delta_1$ could be related empirically to the electric field via $\Delta_1 = \alpha D$ with $\alpha$ = 0.073 e nm$^{-1}$ assuming a linear relationship. Using this relation, the experimental data show excellent agreement with single particle simulations further underlining the findings presented in the main text.

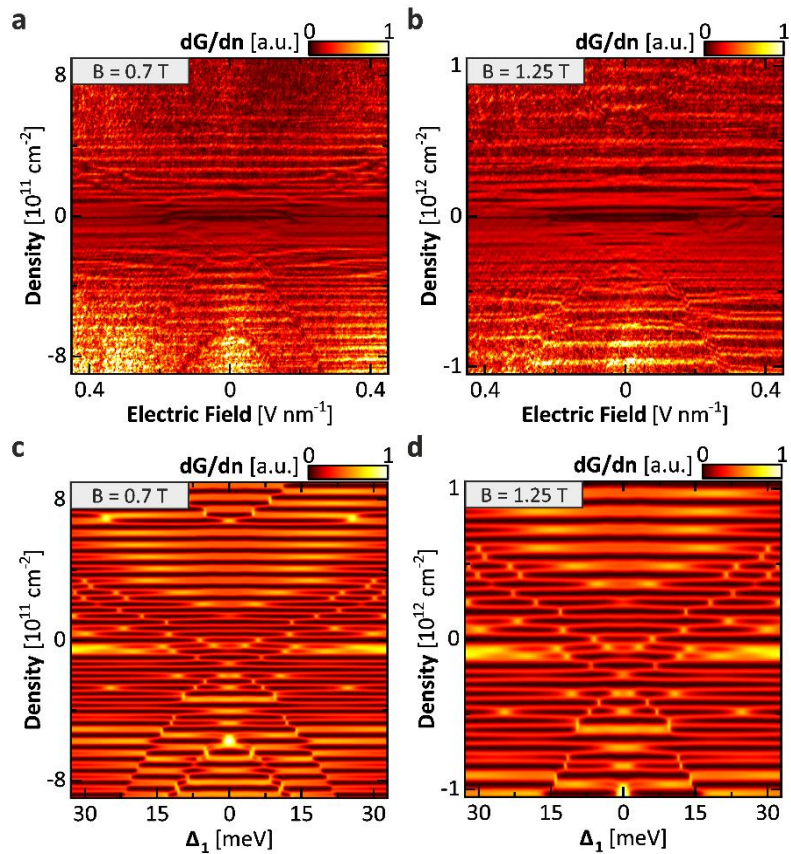

**Figure S3 Magnetotransport at High Charge Carrier Densities: (a,b)** Derivative of the conductance with respect to the charge carrier density as a function of electric field and charge carrier density at a magnetic field of $B$ = 0.7 T (a) and $B$ = 1.25 T (b). **(c,d)** Simulated DOS as a function of charge carrier density and interlayer potential difference at $B$ = 0.7 T (c) and $B$ = 1.25 T (d). The scaling of $\Delta_1$ in (c,d) corresponds to the scaling of $D$ in (a,b).

## Magnetotransport at an Intermediate Electric Field

For completeness, the magnetotransport data at intermediate electric field of D = 0.4 mV nm-1 together with the corresponding simulated evolution of Landau levels are shown in Figure 4. In contrast to the measurements performed at high electric field presented in Figure 3b of the main text, no clear indication of broken-symmetry states is visible even at low magnetic fields.

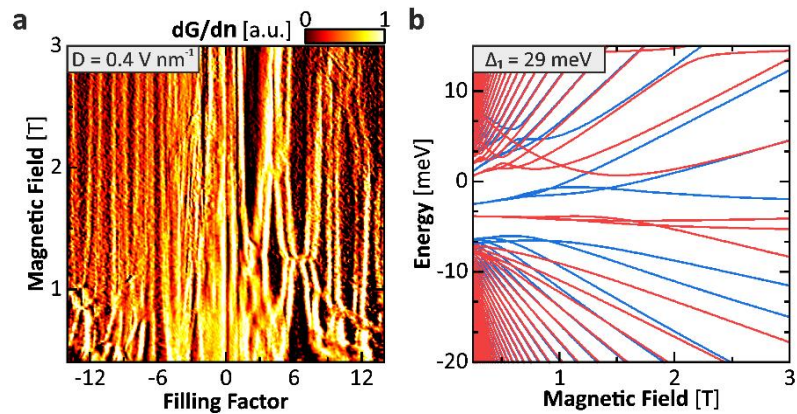

**Figure S4 Magnetotransport at Intermediate Electric Fields: (a)** Derivative of the conductance with respect to filling factor and the magnetic field at an electric field of $D$ = 400 mV nm$^{-1}$. Conductance plateaus are indicated by a vanishing conductance derivative. **(b)** Simulated evolution of spin-degenerate LLs at potential differences of $\Delta_1$ = 29 meV stemming from the $K$ (red) and $K'$ (blue) valley as a function of the magnetic field. The value for $\Delta_1$ corresponds to the electric fields in (a).

## Magnetotransport Zoom-in

In order to highlight the change in LL degeneracy, a zoom-in of Figures 2a,b is provided in Figure S5

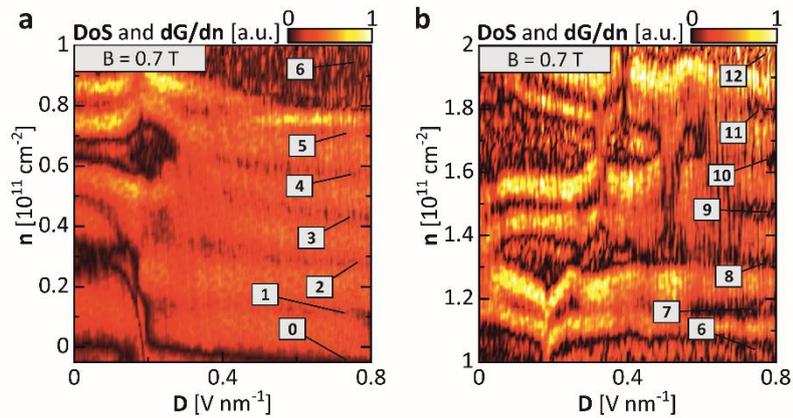

**Figure S5 Landau Levels in Electric Fields:** Zoom-in on the low density (a) and high-density (b) region of Figure 2a in the main text. The contour plots shows the conductance derivative of b-TLG as a function of potential difference $\Delta_1$ and electric field $D$ at $B = 0.7$ T. Dark regions correspond to vanishing DOS or derivative of the conductance (i.e. a conductance plateau) respectively.